\documentstyle[11pt,newpasp,twoside,psfig]{article}
\begin{document}

\title{ Line Profiles from Different Accretion Engine Geometries}
\author{Eric G. Blackman}
\affil{Dept. of Physics and Astron., Univ. of Rochester, Rochester NY 14627}
\author{Sean A. Hartnoll}
\affil{DAMTP, Cambridge Univ., Wilberforce Rd., Cambridge CB3 OWA, UK}

\setcounter{page}{111}
\index{Blackman, E.G.}
\index{Hartnoll, S.A}
\index{flourescent iron lines}
\index{line profiles}
\index{accretion}
\index{black holes}
\index{engine geometry}
\index{X-ray reprocessing}



\section{Motivation and Example Profiles for Four Classes of Geometries}

Cold matter in AGN and some X-ray binary engines 
can reprocess the primary X-ray continuum (e.g. George \& Fabian 1991)
and produce a fluorescent iron K  emission line near 6.4keV whose
profile contains information about the geometry/dynamics 
of the accretion engine (Fabian et al. 2000). 
Thin discs illuminated from above by an X-ray point souce 
reproduce many data features (e.g. Tanaka et al.
 1995; Nandra et al. 1997) but variety in the data still motivates 
calculation of line profiles from other plauible disc geometries. 
(all profiles shown below are computed for Schwarzchild holes):

{\bf Concavity (Fig.~1, left)}:
Sufficiently concave discs 
enhance the peak near the rest frequency (Blackman 1999), 
an effect which may also be transient.

{\bf Warps (Fig.~1, right)}: Here nonaxisymmetry is important
(Hartnoll \& Blackman 2000): the inclination angle and the azimuthal
 viewing angle both determine the profile, unlike
a flat disc for which  all azimuthal viewing angles are equivalent. 
Time variability of the line profile is expected on the
 time-scale of any warp precession around the disc. 
Shadowing also changes the equivalent width of the line as a function of azimuthal viewing angle. Other possible features:
 (1) sharper peaks near the rest frequency compared to flat disc.
 (2) sharp red peaks or steep red fall-offs, and/or 
soft blue fall-offs from shadowing and nonaxisymmetry.
 (3) deep minima states
from shadowing.

{\bf Clumps (Fig.~2, left)}: This alternative, including outflows,
(Hartnoll \& Blackman 2001) is motivated by studies of 
the survival of cold dense clouds in AGN engines (Kuncic et al. 1997). 
When embedded in advection dominated accretion 
flows (ADAFs, e.g. Narayan \& Yi 1995), iron lines can be produced.
These profiles have  less sensitivity to inclination angle than flat discs.

{\bf Spirals (Fig.~2, right)}:
 Line profiles for discs which have
spiral velocity structure (Hartnoll \& Blackman 2002)
show  multiple, quasi-periodic sub-peaks and/or  step-like
 structures in the blue wing of the line profile for high inclination discs, and a dependence on azimuthal viewing angle from nonaxisymmetry. 
Time variability is expected on time scales of the sprial wave propagation.

\bigskip
\begin{figure}[t]
\centerline{\psfig{figure=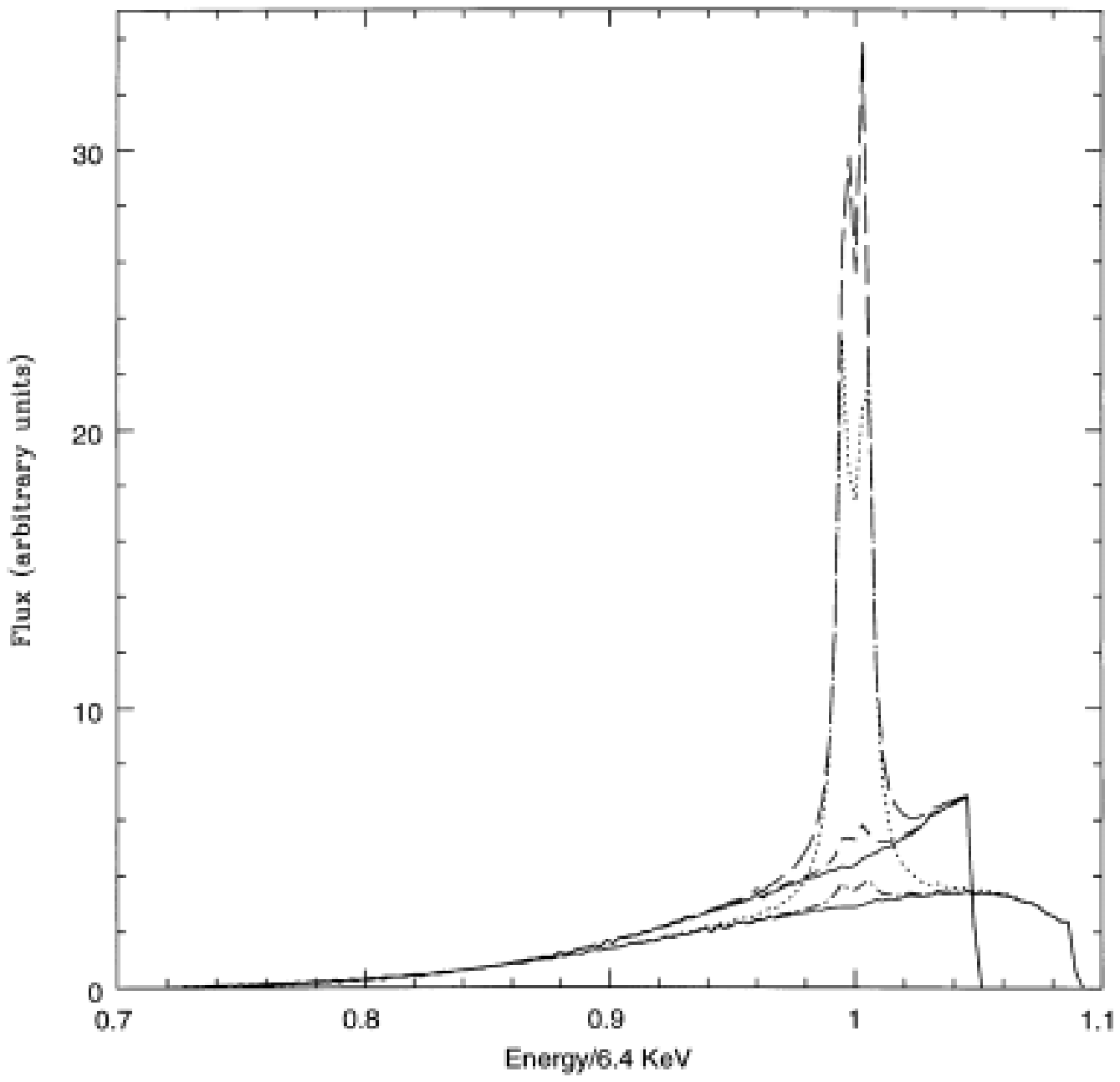,width=5cm}
\psfig{figure=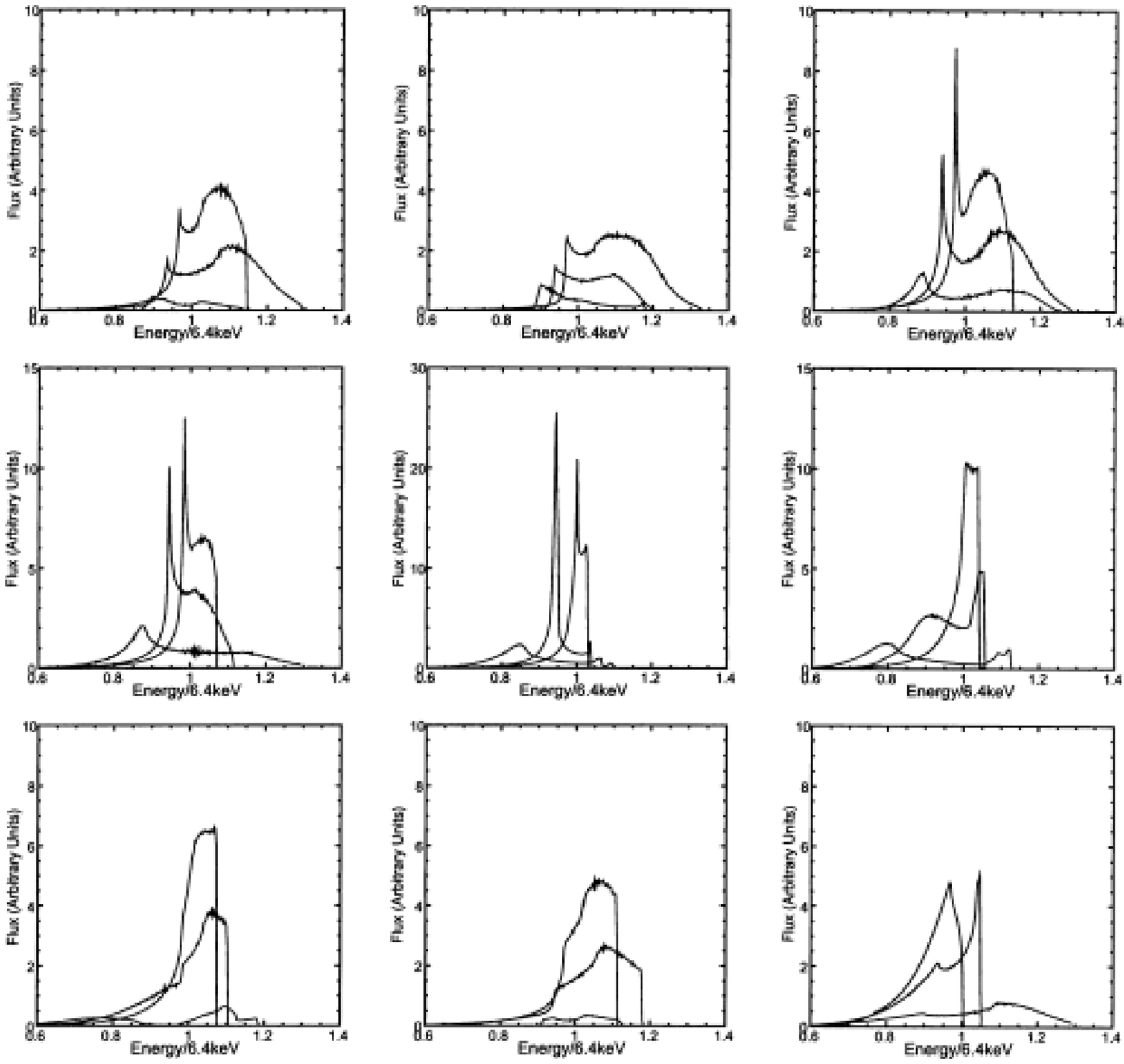,width=5cm}}
\caption{{\it Left box}: Profiles for flat and concave models. 
Solid lines are flat discs at inclinations of 40  
(broad curve) and 30 deg (narrow curve). 
{\it Right boxes}: warped disc profiles; 
top, centre and  bottom lines in each panel represent inclinations  
of 10, 30, 70 deg.
Panels show profiles seen from different azimuths.  
Bottom right is flat disc.}
\label{fig1}
\end{figure}
\bigskip
\begin{figure}[t]
\centerline{\psfig{figure=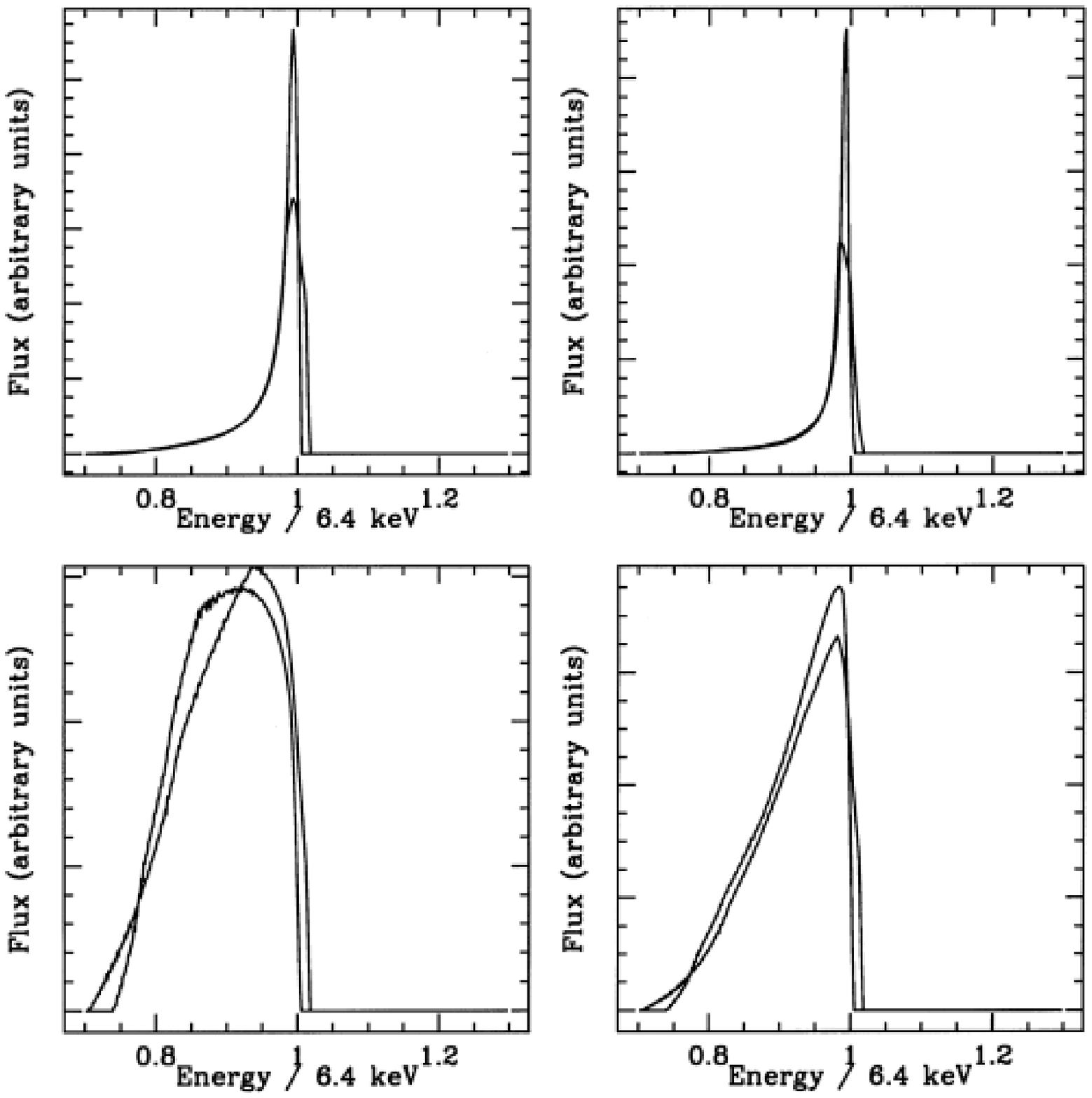,width=4cm}\psfig{figure=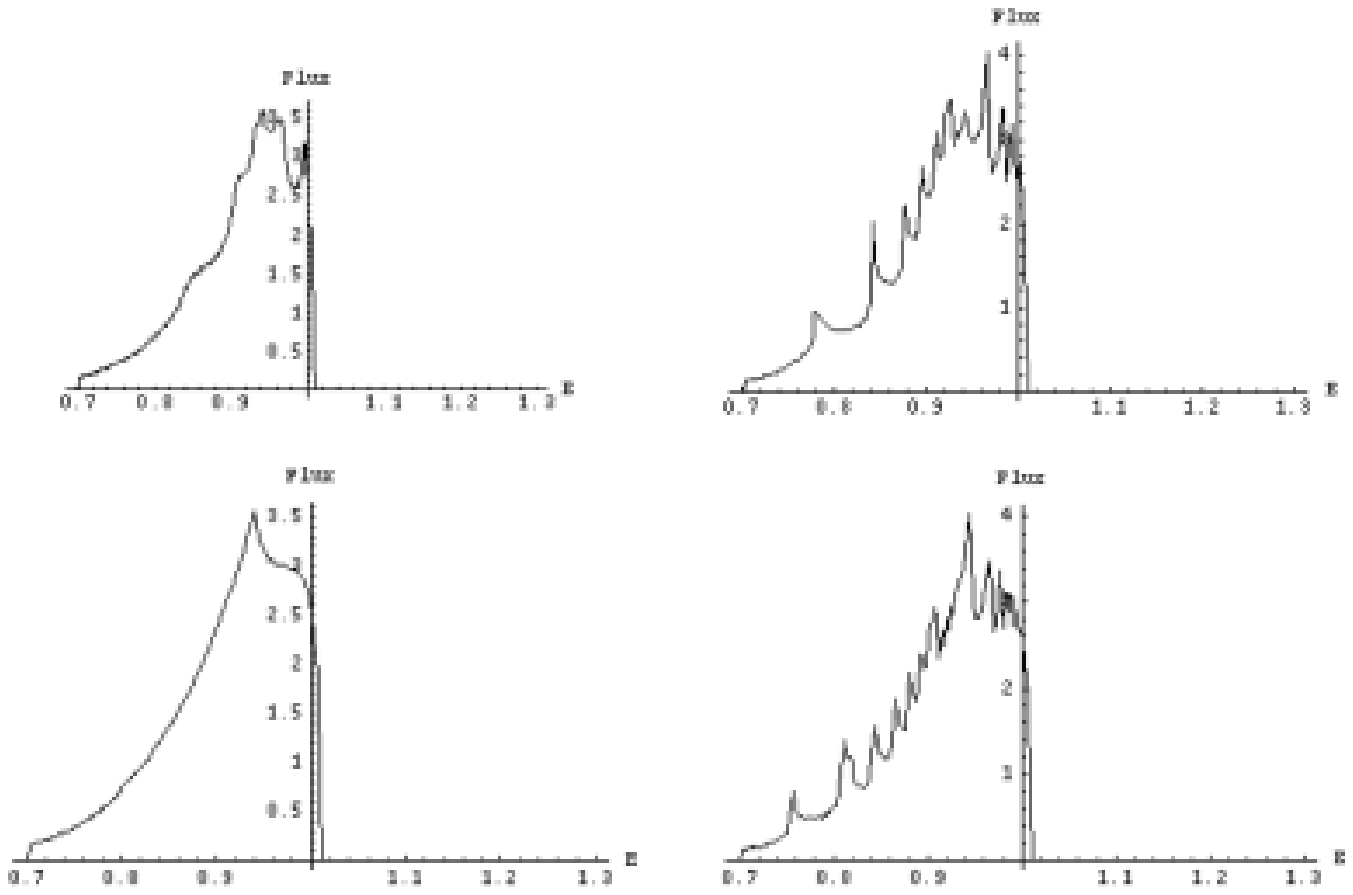,width=6cm}}
\caption{{\it Left 4 boxes}: Clouds in ADAF discs. 
Column 1: little shadowing,
Column 2: signficant shadowing,
top row: outer clouds dominate.
bottom row: inner clouds dominate. Inclinations are 
15 and 60 (higher blue cutoff) deg.
{\it Right 4 boxes}: discs with spirals, 15 deg inclination
and different numbers of spiral arms (0 in bottom left).
}\label{fig3}
\end{figure}

\bigskip
\bigskip




\end{document}